\documentclass[preprint,prd,superscriptaddress,showpacs]{revtex4}
\usepackage{graphicx}
\usepackage{dcolumn}
\usepackage{bm}
\newcommand{\p}{\partial}
\newcommand{\pslash}{p\kern-1ex /}
\newcommand{\lslash}{l\kern-1ex /}
\newcommand{\kslash}{k\kern-1ex /}
\newcommand{\dslash}{\p\kern-1.2ex /}
\newcommand{\Dslash}{{\cal D}\kern-1.5ex /}
\newcommand{\Aslash}{A\kern-1.2ex /}

\newcommand{\tr}{{\rm tr}}

\newcommand{\bea}{\begin{eqnarray}}
\newcommand{\eea}{\end{eqnarray}}
\newcommand{\vol}{\Omega}

\newcommand{\BAN}{\begin{eqnarray*}}
\newcommand{\EAN}{\end{eqnarray*}}
\begin{document}

\preprint{UTHEP-549}
\preprint{NTUTH-07-505E}
\preprint{RIKEN-TH-116}
\preprint{KEK-CP-198}
\preprint{YITP-07-59}
\title{
  Topological susceptibility in two-flavor lattice QCD with exact
  chiral symmetry
}

\newcommand{\NTU}{
  Physics Department, Center for Theoretical Sciences, 
  and National Center for Theoretical Sciences, 
  National Taiwan University, Taipei~10617, Taiwan  
}

\newcommand{\RIKEN}{
  Theoretical Physics Laboratory, RIKEN,
  Wako 351-0198, Japan
}

\newcommand{\Tsukuba}{
  Graduate School of Pure and Applied Sciences, University of Tsukuba,
  Tsukuba 305-8571, Japan
}

\newcommand{\BNL}{
  Riken BNL Research Center, Brookhaven National Laboratory, Upton,
  NY11973, USA
}

\newcommand{\KEK}{
  High Energy Accelerator Research Organization (KEK),
  Tsukuba 305-0801, Japan
}

\newcommand{\GUAS}{
  School of High Energy Accelerator Science,
  The Graduate University for Advanced Studies (Sokendai),
  Tsukuba 305-0801, Japan
}

\newcommand{\YITP}{
  Yukawa Institute for Theoretical Physics, 
  Kyoto University, Kyoto 606-8502, Japan
}

\newcommand{\HUDP}{
  Department of Physics, Hiroshima University,
  Higashi-Hiroshima 739-8526, Japan
}

\newcommand{\RCAS}{
  Research Center for Applied Sciences, Academia Sinica,
  Taipei~115, Taiwan
}

\author{S.~Aoki}
\affiliation{\Tsukuba}
\affiliation{\BNL}

\author{T.W.~Chiu}
\affiliation{\NTU}

\author{H.~Fukaya}
\affiliation{\RIKEN}

\author{S.~Hashimoto}
\affiliation{\KEK}
\affiliation{\GUAS}

\author{T.H.~Hsieh}
\affiliation{\RCAS}

\author{T.~Kaneko}
\affiliation{\KEK}
\affiliation{\GUAS}

\author{H.~Matsufuru}
\affiliation{\KEK}

\author{J.~Noaki}
\affiliation{\KEK}

\author{K.~Ogawa}
\affiliation{\NTU}

\author{T.~Onogi}
\affiliation{\YITP}

\author{N.~Yamada}
\affiliation{\KEK}
\affiliation{\GUAS}


\collaboration{JLQCD and TWQCD Collaborations}
\noaffiliation

\pacs{11.15.Ha,11.30.Rd,12.38.Gc}

\begin{abstract}
  We determine the topological susceptibility $\chi_t$ in two-flavor QCD using
  the lattice simulations at a fixed topological sector.
  The topological charge density is unambiguously defined on the lattice using
  the overlap-Dirac operator which 
  possesses exact chiral symmetry.
  Simulations are performed on a $16^3 \times 32$ lattice at lattice  
  spacing $\sim$ 0.12~fm at six sea quark masses $m_q$ ranging in
  $m_s/6$--$m_s$ with $m_s$ the physical strange quark mass.
  The $\chi_t$ is extracted from the constant behavior of 
  the time-correlation of flavor-singlet pseudo-scalar meson 
  two-point function at large distances,
  which arises from the finite size effect due to the fixed topology.  
  In the small $m_q$ regime, our result of $\chi_t$ is proportional to $m_q$
  as expected from chiral effective theory. 
  Using the formula $\chi_t=m_q\Sigma/N_f$ by Leutwyler-Smilga, we obtain the
  chiral condensate in $N_f=2$ QCD as 
  $\Sigma^{\overline{\mathrm{MS}}}(\mathrm{2~GeV}) = 
  [252(5)(10) \mathrm{MeV}]^3 $, in good agreement with
  our previous result obtained in the $\epsilon$-regime. 
\end{abstract}

\maketitle

The vacuum of Quantum Chromodynamics (QCD) has a non-trivial 
topological structure.
The cluster property and the gauge invariance require that the ground state
must be the $\theta$ vacuum, a superposition of gauge configurations in
different topological sectors.
The topological structure is also essential for the $U(1)$ problem, 
{\it i.e.} the flavor singlet pseudo-scalar meson {\it knows} 
the presence of non-trivial topological charge in the QCD vacuum.
Therefore, the topological excitations, such as the instantons, played a
central role in the past attempts to understand the QCD vacuum since the
early years of QCD.

The topological charge fluctuations are characterized by the topological
susceptibility $\chi_t$, defined as 
$\chi_t\equiv\left< Q^2\right>/\Omega$ through the topological charge $Q$ in a
given volume $\vol$.
Since the topological excitations do not occur in the perturbation theory,
theoretical calculation of $\chi_t$ from the QCD Lagrangian necessarily
involves non-perturbative methods, such as the numerical simulation of lattice
QCD.  

Lattice calculation of $\chi_t$ has become reasonably precise in the quenched
approximation \cite{Del Debbio:2004ns,Durr:2006ky}, which is however not
realistic as the quark loop effects are discarded. 
On the other hand, despite much effort, 
more realistic calculation without
quenching remains a challenging problem. 
For instance, in full QCD, $\chi_t$ is expected to linearly depend on the
quark mass and to vanish in the chiral limit. 
Even such a basic property has not been clearly reproduced in the previous
lattice calculations (for a recent study, see, e.g., \cite{Bernard:2003gq}). 

The main difficulties are the following. 
(i) Definition of the topological charge density using the gauge links
(discretization of 
$(1/32\pi^2)\epsilon_{\mu\nu\rho\sigma} \tr(F_{\mu\nu}{F}_{\rho\sigma})$)
causes bad ultraviolet divergences, while  
the cooling methods devised to tame the short distance fluctuations 
introduce sizable systematic uncertainties. 
(ii) Unquenched simulations with Wilson/staggered fermion do not respect
correct chiral or flavor symmetry at finite lattice spacing, and the
definition of the topological charge through the Atiyah-Singer index theorem
is ambiguous.
(iii) With the molecular-dynamics-type algorithms, which are based on a
continuous evolution of the gauge links, the system is trapped in a fixed
topological sector as the continuum limit is approached.
Therefore, a proper sampling of different topological sectors cannot be
achieved.  
Approaching the chiral limit (or the physical up and down quark masses) also
makes the tunneling of topological charge a rare event, because of the
suppression of the fermion determinant for large topological charges.

During the last decade, (i) and (ii) have been solved by the realization of
exact chiral symmetry on the lattice, with which the topological charge is
uniquely defined at any finite lattice spacing by counting the number of
fermionic zero-modes.
(For recent studies respecting the exact chiral symmetry, we refer
\cite{DeGrand:2005vb,Egri:2005cx}.)
However, (iii) remains insurmountable, since the correct sampling of topology
becomes increasingly more difficult towards realistic simulation with lighter
quarks and finer lattices.
A plausible solution is to perform QCD simulations in a fixed topological
sector and to extract $\chi_t$ from local topological fluctuations.
In \cite{Aoki:2007ka} (see also \cite{Brower:2003yx}), a general formula to
transcribe any observable measured at a fixed topological charge to its value
in the $\theta$ vacuum is derived.
As an application, a new method to calculate $\chi_t$ is also proposed.

In this paper, we use this method to precisely calculate $\chi_t$ in
two-flavor lattice QCD with exact chiral symmetry. 
The results are compared with the prediction from chiral perturbation theory
($\chi$PT) \cite{Leutwyler:1992yt}
\begin{equation}
  \label{eq:LS}
  \chi_t = \frac{m_q \Sigma}{N_f} + {\cal O}(m_q^2),
\end{equation}
for $N_f$ flavors of sea quarks with mass $m_q$.
The chiral condensate $\Sigma$ can be determined independently, {\it e.g.}
from the low-lying eigenvalues of the Dirac operator
\cite{Fukaya:2007fb,Fukaya:2007yv}.
Therefore, the comparison provides a critical test of the lattice approach to
study the QCD vacuum in the chiral regime.

A two-point function of the topological charge density $\rho(x)$
calculated in a finite volume $\vol$ at a given topological charge $Q$ behaves
as \cite{Aoki:2007ka}
\begin{equation}
  \label{eq:rho_rho}
  \lim_{|x| \to \infty} \left< \rho(x) \rho(0) \right>_Q =
  \frac{1}{\vol} \left( \frac{Q^2}{\vol} - \chi_t 
                       -\frac{c_4}{2 \chi_t \vol} \right)
   + {\cal O}(\vol^{-3})
\end{equation}
where 
$c_4 = -(   \langle Q^4 \rangle
         -3 \langle Q^2 \rangle^2 )/\vol $.
The expectation value $\left<\cdots\right>_Q$ denotes an average in a given
topological sector $Q$. 
The correlation does not vanish even for large separations, because of the
violation of the clustering property at fixed topological charge.
We emphasize that the derivation of (\ref{eq:rho_rho}) relies only on modest
assumptions such as $\left< Q^2\right>\gg 1$ and $Q\ll \left< Q^2\right>$,
which are the conditions to apply the saddle point expansion in the Fourier
transform from a fixed $\theta$ to a fixed $Q$.
Except for these conditions, the formula is model independent.

We consider, in particular, two spatial sub-volumes at $t_1$ and $t_2$,
for which the correlator is defined as
\begin{equation}
\label{eq:Ct}
  C(t_1 - t_2) \equiv \langle Q(t_1) Q(t_2) \rangle
  = \sum_{\vec{x}_1,\vec{x}_2} \left<\rho(x_1) \rho(x_2) \right>,
\end{equation}
where the summations run over the spatial sites $\vec{x}_1$ and $\vec{x}_2$ at
$t_1$ and $t_2$, respectively. 
Its plateau at large $|t_1-t_2|$ can be used to extract $\chi_t$, provided
that $|c_4|\ll 2 \chi_t^2\vol$. 

In order to preserve the exact chiral symmetry, which is essential for the
definition of the topological charge, we employ the
overlap-Dirac operator \cite{Neuberger:1997fp,Narayanan:1995gw}
\begin{equation}
  \label{eq:overlap}
  D(m_q)=\left(m_0+\frac{m_q}{2}\right)
  + \left(m_0-\frac{m_q}{2}\right)\gamma_5 \mathrm{sgn}\left[H_W(-m_0)\right]
\end{equation}
with mass $m_q$.
The kernel operator $H_W(-m_0)$ is the conventional Wilson-Dirac operator with
a large negative mass term $-m_0$.

\begin{figure}[tb]
  \centering
  \includegraphics[width=8cm,clip=true]{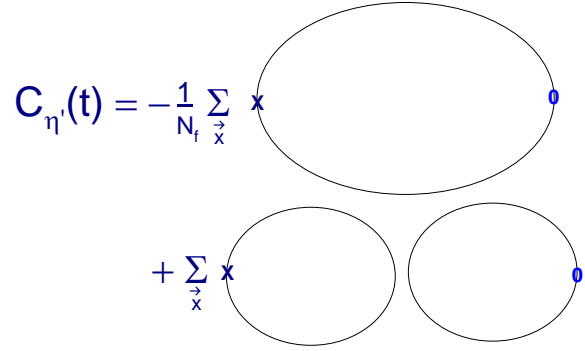}
  \caption{
   A schematic diagram for the time-correlation function 
   of the flavor singlet operator $P^0(x)$. 
   Each solid line denotes the valence quark propagator.
  }
\label{fig:etap}
\end{figure}

In place of the topological charge density $\rho(x)$ (and $\rho(0)$) 
in (\ref{eq:rho_rho}), we use $ m_q P^0(x) $ (and $m_q P^0(0)$), 
that were shown to give the same asymptotic 
constant as (\ref{eq:rho_rho}) \cite{Aoki:2007ka} 
(the original suggestion is in \cite{Fukaya:2004kp}), where $ P^0 (x) $ 
is the flavor singlet pseudo-scalar density
$ P^0(x)\equiv\frac{1}{N_f}\sum_{f=1}^{N_f}\bar{\psi}^f(x)
                   \gamma_5[1-aD(0)/(2m_0)]\psi^f(x)
$.
The correlator
$C_{\eta'}(t)\equiv\sum_{\vec{x}}\left<P^0(x)P^0(0)\right>$
contains a connected and a disconnected diagram
as shown in Fig.~\ref{fig:etap}.
If we pick the disconnected piece and identify a
``topological charge density'', it can be written as
$ \rho_1(x) = m_q \mathrm{tr}[\gamma_5(D_c+m_q)^{-1}_{x,x}] $, 
where $D_c$ is a chirally-symmetric ($\gamma_5D_c+D_c\gamma_5=0$)
nonlocal operator, relating to
$D(0)$ by $D_c=[1-aD(0)/(2m_0)]^{-1}D(0)$ \cite{Chiu:1998gp}.
Integrated over the entire lattice volume, $\rho_1(x)$ 
reduces to the number of fermionic zero-modes, and thus
has the necessary property for the topological charge
density. This implies that the correlator $\left<\rho_1(x)\rho_1(0)\right>$ 
has the same asymptotic constant as (\ref{eq:rho_rho}). 
However, the correlator $\left<m_qP^0(x)m_qP^0(0)\right>$ 
approaches the constant with the rate governed by the $\eta'$ mass,
$e^{-m_{\eta'}|x|}$, which is much faster than
$e^{-m_{\pi}|x|}$ appearing in $\left<\rho_1(x)\rho_1(0)\right>$.

Simulations are carried out for two-flavor ($N_f=2$) QCD on a $16^3\times 32$
lattice at a lattice spacing $\sim$ 0.12~fm. 
For the gluon part, the Iwasaki action is used at $\beta$ = 2.30 together with 
unphysical Wilson fermions and associated twisted-mass ghosts
\cite{Fukaya:2006vs}.
The unphysical degrees of freedom generate a factor 
$\det[H_W^2(-m_0)/(H_W^2(-m_0)+\mu^2)]$ in the partition function 
(we take $m_0=1.6$ and $\mu=0.2$)
that suppresses the near-zero eigenvalue of $H_W(-m_0)$ and thus
makes the numerical operation with the overlap operator (\ref{eq:overlap}) 
substantially faster. 
Furthermore, since the exact zero eigenvalue is forbidden, the global
topological change is preserved during the molecular dynamics evolution of the
gauge field.  Our main runs are performed at $Q=0$, while $Q=-2$ and
$-4$ configurations are also generated at one sea quark
mass in order to check the consistency as described below.
Ergordicity within a given global topological charge is satisfied if the
configuration space of that topological sector forms a connected manifold.
This is indeed the case in the continuum SU(3) gauge theory on a
four-dimensional torus, and therefore is probably also true at small lattice
spacing adopted in this work.

We use the Hybrid Monte Carlo algorithm \cite{Duane:1987de} with the mass
preconditioning \cite{Hasenbusch:2001ne}.
The fermion masses for the preconditioner were chosen to be 0.4 for 
heavier sea quark masses and 0.2 for the two lightest ones (see later).
We exploit the rational approximation {\it a la} Zolotarev for the sign
function in (\ref{eq:overlap}) after projecting out low-lying eigenmodes of
$H_W(-m_0)$. With the number of poles in the rational function to be 8--10, 
the accuracy of $O(10^{-(7-8)})$ is achieved for the sign function.
The simulations have been done in two phases for each sea quark mass. 
In the first phase the nested conjugate gradient (CG) 
is used to invert the overlap operator (\ref{eq:overlap}) 
(see \cite{Kaneko:2006pa,Aoki:2008tq} for details).
On the other hand, in the second phase we use the five-dimensional
implementation of the overlap solver without the low-mode projection.
The target accuracy of $O(10^{-(7-8)})$ is maintained by adding an additional
Metropolis step calculated with the nested CG \cite{Matsufuru:2006xr}.

For the sea quark mass $m_q$ we take six values: 0.015, 0.025, 0.035, 0.050,
0.070, and 0.100 that cover the mass range $m_s/6$--$m_s$ with $m_s$ the
physical strange quark mass.
After discarding 500 trajectories for thermalization, we accumulate 
10,000 trajectories in total for each sea quark mass.
In the calculation of $\chi_t$, we take one configuration every 20
trajectories, thus we have 500 configurations for each $ m_q $.
For each configuration, 50 conjugate pairs of lowest-lying eigenmodes 
of the overlap-Dirac operator $D(0)$ are calculated using the
implicitly restarted Lanczos algorithm and stored for the later use.
In these calculations, the better accuracy of $O(10^{-12})$ is enforced for the
sign function by increasing the number of poles in the rational approximation.

For the connected diagram
(see Fig.~\ref{fig:etap}), 
the pion correlator is computed using the conjugate gradient algorithm 
with a low-mode preconditioning.
Low-modes are also used for averaging over source points
\cite{DeGrand:2004qw}, which significantly improves the statistical signal.
For the disconnected diagram, the quark propagator is represented by the
eigenmode decomposition and approximated by the 50 conjugate pairs of the 
low-lying eigenmodes.
The quark propagator is then obtained for any source point without extra
computational cost, and the disconnected loops can be calculated with an
average over the source point.
The truncation is motivated by the expectation that the long distance
correlation is dominated by the low-lying fermion modes; its validity has
to be checked numerically (see below).

In Fig.~\ref{fig:etap_m0025}, we plot $C_{\eta'}(t)$ together with those of
connected and disconnected parts for $m_q = 0.025$.
The curve is the fit to the function $A+B(e^{-Mt}+e^{-M(T-t)})$ with data for
$C_{\eta'}(t)$ in the range $t\in [4,28]$. 
The horizontal line is the fitted constant $A$, and the curvature 
of the data points represents the contamination by the flavor-singlet 
state $\eta'$, that rapidly decays due to its heavy mass.
Assuming $|c_4|\ll 2\chi_t^2\vol$, we obtain 
$a^4 \chi_t = 3.40(27) \times 10^{-5}$ at $m_q = 0.025$.
The error is estimated using the jackknife method with bin size of
20 configurations, with which the statistical error saturates.

\begin{figure}[tb]
  \centering
  \includegraphics[width=9cm,clip=true]{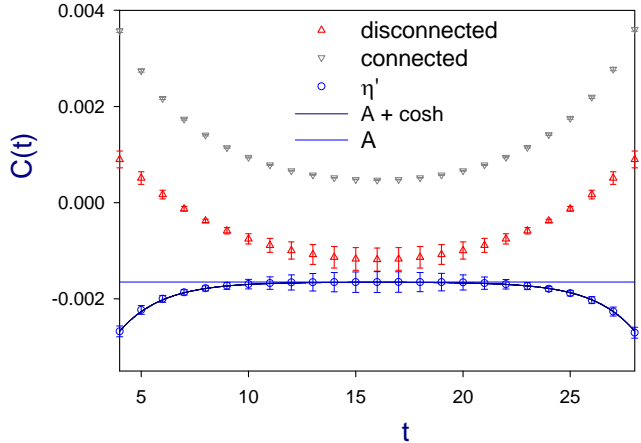}
  \caption{
    Time-correlation function of the flavor singlet $\eta'$ (circles) 
    and its connected (triangle down) and disconnected (triangle up)
    contributions.
    Data at $m_q=0.025$ are shown.
  }
  \label{fig:etap_m0025}
\end{figure}

Since the disconnected diagram is computed with only 50 pairs of low-lying 
eigenmodes, we have to check whether they suffice to saturate $C_{\eta'}(t)$.
For the time range $[4,28]$ used for fitting, 
as the number of eigenmodes is increased from 10 to 30, the change of
correlator $|\delta C_{\eta'}|/C_{\eta'}$ is $\sim 3\%$, while from 30 to 50, 
it is only $\sim 0.3\%$, which is less than 8\% of the statistical error.
Thus $C_{\eta'}$ is well saturated with 50 eigenmodes. 
This also holds for all six sea quark masses.

\begin{figure}[tb]
  \centering
  \includegraphics[width=8cm,clip=true]{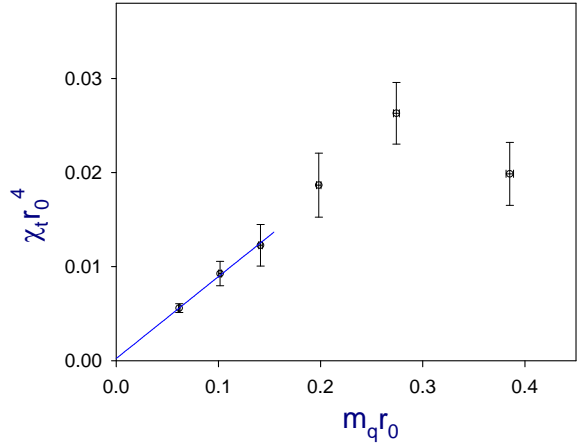}
  \caption{
    Topological susceptibility $ \chi_t r_0^4 $ 
    versus sea quark mass $m_q r_0$.}
  \label{fig:chit_mq_Q0}
\end{figure}

In Fig.~\ref{fig:chit_mq_Q0}, we plot the topological susceptibility 
$ \chi_t r_0^4 $ as a function of the sea quark mass $m_q r_0 $,  
where $ r_0 $ is the Sommer scale extracted from the static quark
potential.    
We also present our data in Table~\ref{tab:chit_Q0}, together with
the lattice spacing $ a $ determined from the static quark potential 
with the input $ r_0 = 0.49 $ fm \cite{Aoki:2008tq}, 
and the pseudoscalar mass $ m_\pi r_0 $ \cite{Noaki:2007es}.
The statistical precision is good enough to find a clear dependence 
on the sea quark mass.
For the smallest three quark masses, 0.015, 0.025, and 0.035, the data are
well fitted by a linear function $F+G m$ with the intercept 
$F = 0.2(3) \times 10^{-3} $ and the slope $G = 0.087(4)$.
Evidently, the intercept is consistent with zero, in agreement 
with the $\chi$PT expectation (\ref{eq:LS}). 
Equating the slope to $r_0^3 \Sigma/N_f$, we obtain 
$r_0^3\Sigma = 0.174(8)$.

In order to convert $\Sigma$ to that in the
$\overline{\mathrm{MS}}$ scheme, we calculate the 
renormalization factor $Z_m^{\overline{\mathrm{MS}}}(\mathrm{2~GeV})$
using the non-perturbative renormalization technique
through the RI/MOM scheme \cite{Martinelli:1994ty}.
Our result is 
$Z_m^{\overline{{\text MS}}}(\mbox{2 GeV}) = 0.742(12)$ 
\cite{Noaki:2007es}
With $ a^{-1} = 1670(20)(20)$ MeV determined with $ r_0 = 0.49 $ fm 
\cite{Kaneko:2006pa,Aoki:2008tq}, 
the value of $ \Sigma $ is transcribed to
$ \Sigma^{\overline{{\text MS}}}(\mbox{2 GeV}) = [252(5)(10) \mbox{ MeV}]^3 $, 
which is in good agreement with our previous result
$ [251(7)(11) \mbox{ MeV}]^3 $ \cite{Fukaya:2007fb,Fukaya:2007yv}
obtained in the $\epsilon$-regime from the low-lying eigenvalues.
The errors represent a combined statistical error
($a^{-1}$ and $Z_m^{\overline{\mathrm{MS}}}$) and
the systematic error estimated from the higher order effects 
({\it e.g.}, $ c_4 $ term), respectively.
Since the calculation is done at a single lattice spacing,
the discretization error cannot be quantified reliably, but
we do not expect much larger error because our lattice
action is free from $O(a)$ discretization effects.

\begin{table}
\begin{center}
\caption{The topological susceptibility $ \chi_t r_0^4 $ versus the quark mass
         $m_q r_0 $ obtained in this work. The lattice spacing $ a $
         for each quark mass is determined from the static quark potential
         with the input $ r_0 = 0.49 $ fm \cite{Aoki:2008tq}.}
\vspace{0.5cm}
\begin{tabular}{c|c|c|c|c|c}
\hline
$ m_q a $ & $a$[fm] & $ m_\pi r_0 $ & $ m_q r_0 $ & $\chi_t r_0^4 $ 
          & $ \chi_t \vol $ \\
\hline
\hline
0.015 & 0.1194(15) & 0.7096(102) &  0.0616(8) 
      & $ 5.59(47) \times 10^{-3} $  & 2.582(88) \\
0.025 & 0.1206(18) & 0.8935(137) &  0.1016(15) 
      & $ 9.26(1.29) \times 10^{-3} $ & 4.454(356) \\
0.035 & 0.1215(15) & 1.053(13) &  0.1412(17) 
      & $ 0.0123(22) $ &  6.078(794)  \\
0.050 & 0.1236(14) & 1.238(14) &  0.1982(22) 
      & $ 0.0187(34) $ &  9.900(1.354)\\
0.070 & 0.1251(13) & 1.456(15) &  0.2742(28) 
      & $ 0.0263(33) $ & 14.65(1.21) \\
0.100 & 0.1272(12) & 1.734(17) &  0.3852(36) 
      & $ 0.0199(33) $ &  11.82(1.55) \\
\hline
\end{tabular}
\label{tab:chit_Q0}
\end{center}
\end{table}

In principle, $\chi_t$ in (\ref{eq:rho_rho}) is universal for any 
topological sector.
We check the universality of $ \chi_t $ as follows. 
At sea quark mass $m_q=0.050$, we generate 250 configurations with $Q=-2$
and $-4$ respectively in addition to the main run at $Q=0$.  
Then we extract $\chi_t$ from the time-correlation function of $\eta'$, 
similar to the $Q=0$ case. 
Our results for $a^4 \chi_t$ are: 
$\{ 7.4(1.3), 6.4(2.1), 5.9(1.8) \} \times 10^{-5}$
for $Q=\{0,-2,-4\}$ respectively. 
Evidently, $\chi_t$ extracted from different topological sectors 
are consistent with each other within the statistical error.

Finally, we discuss on the neglected terms in (\ref{eq:rho_rho}).
The leading correction comes from the $|c_4|\ll 2\chi_t^2\vol$ term.
With the formulas derived in \cite{Aoki:2007ka}, we can obtain 
an estimate of the upper bound of $|c_4|/(2\chi_t^2 \vol)$ by 
measuring the two-point correlator $\langle\rho_1(x_1)\rho_1(x_2)\rangle$ and
the four-point correlator
$\langle\rho_1(x_1)\rho_1(x_2)\rho_1(x_3)\rho_1(x_4)\rangle$. 
Our preliminary result is $ |c_4|/(2 \chi_t^2 \vol) < 0.1 $,  
for all six sea quark masses.
Details of this calculation will be presented elsewhere.
We note that this upper bound 
is also consistent with the ratio 
$|c_4|/\chi_t\simeq 0.3$ obtained in the quenched approximation  
\cite{Giusti:2007tu,Del Debbio:2007kz}.
  
In this letter, we have determined the topological susceptibility $\chi_t$
in two-flavor QCD from a lattice calculation of two-point correlators at a
fixed global topological charge.
The expected sea quark mass dependence of $\chi_t$ from $\chi$PT is clearly
observed with the good statistical precision we achieved, in contrast to the
previous unquenched lattice calculations.
Our result indicates that the topologically non-trivial excitations 
({\it e.g.}, instanton and anti-instanton pairs) 
are in fact locally active in the QCD vacuum, even when the global topological
charge is kept fixed.
The information of these topological excitations is carried by low-lying
fermion eigenmodes if the exact chiral symmetry is respected on the lattice.
This work demonstrates that Monte Carlo simulation of lattice QCD with 
fixed topology is a viable approach, to be pursued when the topology change
hardly occurs near the continuum limit even with chirally non-symmetric
lattice fermions.
The artifacts due to the fixed topology in a finite volume can be removed to
obtain the physics results in the $\theta$ vacuum, provided that $\chi_t$ has
been determined in the first place \cite{Aoki:2007ka,Brower:2003yx} as has
been done in this work. 

\begin{acknowledgments}
  Numerical simulations are performed on Hitachi SR11000 and IBM System Blue
  Gene Solution at High Energy Accelerator Research Organization (KEK) under 
  a support of its Large Scale Simulation Program (No.~07-16), and also 
  in part on NEC SX-8 at YITP (Kyoto U), NEC SX-8 at RCNP (Osaka U), 
  and IBM/HP clusters at NCHC and NTU-CC (Taiwan).
  This work is supported in part by the Grant-in-Aid of the
  Japanese Ministry of Education 
  (No.~13135204, 
       15540251,      
       17740171,       
       18034011,      
       18340075,     
       18740167,      
       18840045,      
       19540286,      
  and  19740160)  
  and the National Science Council of Taiwan 
  (No.~NSC96-2112-M-002-020-MY3,
       NSC96-2112-M-001-017-MY3).
  The authors would like to acknowledge YITP workshop YITP-W-05-25, 
  where this work was initiated.
\end{acknowledgments}

\end{document}